\documentclass[useAMS,usenatbib]{mn2e}
\usepackage{graphicx,color,epsfig,amssymb}

\newcommand{\be}{\begin{equation}}
\newcommand{\ee}{\end{equation}}

\newcommand{\apj}{ApJ}
\newcommand{\apjs}{ApJS}
\newcommand{\mnras}{MNRAS}
\newcommand{\aap}{A\&A}
\newcommand{\araa}{ARA\&A}
\newcommand{\apjl}{ApJL}

\newcommand{\icarus}{ICARUS}

\def\ltsima{$\; \buildrel < \over \sim \;$}
\def\simlt{\lower.5ex\hbox{\ltsima}}
\def\gtsima{$\; \buildrel > \over \sim \;$}
\def\simgt{\lower.5ex\hbox{\gtsima}}

\newcommand\mearth{{\,{\rm M}_{\oplus}}}

\def\del#1{{}}

\usepackage{ulem}

\title[Core-assisted gas capture instability]{Core-assisted gas capture
  instability: a new mode of giant planet formation by gravitationally
  unstable discs}

\author[Nayakshin, Helled and Boley]{Sergei Nayakshin$^{1}$, Ravit Helled$^2$
  and Aaron C.~Boley$^{3}$
\\ 
$^{1}$ Department of Physics \& Astronomy, University of Leicester, Leicester, LE1
  7RH, UK. {E-mail:~} {\rm Sergei.Nayakshin@le.ac.uk} \\
$^2$ Department of Geophysics, Atmospheric, and Planetary Sciences, Tel-Aviv
  University, Israel.\\
$^3$ The University of British Columbia, Department of Physics and Astronomy, 6224 Agricultural Rd., Vancouver, BC V6T 1Z1, Canada.}

\begin{document}

\date{Received}

\pagerange{\pageref{firstpage}--\pageref{lastpage}} \pubyear{2008}

\maketitle

\label{firstpage}

\begin{abstract}
Giant planet formation in the core accretion plus gas capture (CA) paradigm is
predicated by the formation of a core, assembled by the coagulation of grains
and later by planetesimals within a protoplanetary disc. As the core mass
increases beyond a critical value, the hydrogen-dominated atmosphere around
the core becomes self-gravitating and collapses onto the core, triggering
rapid gas accretion which can lead to the formation of a gaseous planet.  
  In contrast, in the disc instability paradigm, giant planet formation is
  believed to be independent of core formation: massive self-gravitating gas
  fragments cool radiatively and collapse as a whole independently of whether
  there is a core.

In this paper we show that giant planet formation in the disc instability
model may be also enhanced by core formation for reasons physically very
similar to the CA paradigm. In the model explored here, efficient grain
sedimentation within an initial fragment (rather than the disc) leads to the
formation of a core composed of heavy elements.  We find that massive
atmospheres form around cores and undergo collapse as a critical core mass is
exceeded, analogous to CA theory. The critical mass of the core to initiate
such a collapse depends on the fragment mass and metallicity, as well as core
luminosity, but ranges from less than 1 to as much as $\sim80$ Earth
masses. We therefore suggest that there are two channels for the collapse of a
gaseous fragment to planetary scales within the disc instability model: (i)
H$_2$ dissociative collapse of the entire gaseous clump, and (ii)
core-assisted gas capture, as presented here.  We suggest that the first of
these two is favoured in metal-poor environments and for fragments $\simgt
5-10$ Jupiter masses, whereas the second is favored in metal-rich environments
and fragments of lower mass.
\end{abstract}


\section{Introduction}\label{intro}

The self-gravity of a massive protoplanetary disc may, under certain
conditions \citep{Toomre64}, cause small perturbations within the disc to grow
in amplitude (i.e., a gravitational instability; GI) until the instabilities
saturate in the nonlinear regime. GIs can manifest themselves as spiral arms
and shocks, heat the disc, and drive angular momentum and mass transport
\citep[see][for a review]{DurisenEtal07}, as well as create molecular
abundance variations that could be used as observational telltales of active
GIs (Ilee et al.~2010, Douglas et al. 2013).  If the combination of disc
heating and mass transport fail to saturate (self-regulate) GIs
\citep[e.g.,][]{Gammie01,CossinsEtal09}, then the densest regions of spiral
arms can collapse to form self-bound gaseous clumps, which may be precursors
of giant planets \citep[e.g.,][]{Kuiper51b,Boss98}.

When such fragments first form, their mass is of order the mass of a gas giant
planet \citep[e.g.,][]{BoleyEtal10,Nayakshin10a,ForganRice11}. However, their
internal properties are very far from present day gas giants. In particular,
the fragments are $\sim$ 10-12 orders of magnitude less dense than Jupiter(!),
dominated in mass by molecular hydrogen, have initial central temperatures of
just a few hundred Kelvin, and do not have high-metallicity cores
\citep[although they do contain an admixture of heavy elements and grains,
  perhaps significantly enriched over nebular abundances, see,
  e.g.,][]{BoleyEtal11a}. Removal of gas by tides
\citep{BoleyEtal10,Nayakshin10c} or the accretion of gas
\citep{SW08,KratterEtal10} and planetesimals \citep{HelledEtal06}, plus
nonlinear planet-disc coupling through radiative heat exchange
\citep{VazanHelled12,NayakshinCha13} together make it extremely difficult to
say what becomes of the remaining object. The list of possibilities goes from
rocky and giant planets to brown dwarfs and even low mass stars
\citep{SW08,KratterEtal10,ZhuEtal12a,ForganRice13}.

Rapid inward migration of first fragments
\citep{VB05,VB06,ChaNayakshin11a,BaruteauEtal11,MichaelEtal11,ZhuEtal12a} is a
particular concern for the survivability of giant planets formed by GIs.
Initial fragment sizes are comparable to the corresponding Hill sphere of the
clumps themselves, to within a factor of two or three \citep[see,
  e.g.,][]{BoleyEtal10}. To survive the rapid radial migration, the clumps
need to cool and contract rapidly, so that central temperatures reach $\sim
2000$ K \citep[e.g.,][]{HelledEtal06} {\it before} tidal forces from the star
disrupt the planet. At $T\sim 2000$~K, molecular hydrogen can dissociate,
redirecting energy from pressure support into internal molecular
processes. The resulting dynamical collapse of the protoplanet from sizes of
$\sim 1000$ R$_J$ to a few to tens of R$_J$ \citep{BodenheimerEtal80} is
analogous to the end of the first core stage in star formation
\citep[e.g.,][]{Larson69}, and may allow the protoplanet to avoid tidal
destruction \citep[e.g.,][]{Nayakshin10c}.

In this paper we describe a second, so far insufficiently explored, channel
for the collapse of giant planets faced with rapid inward migration. This
second channel exists entirely due to the heavy element component within the
fragments. As is well known, rapid grain growth can promote sedimentation of
large grains to the centre of the clump
\citep{McCreaWilliams65,Boss98,HelledEtal08}. In this paper we show that {\it
  if} the segregation of heavy elements to the central regions of the fragment
is efficient, then dynamical collapse of the clump may be initiated by a
dynamical instability in the clump's centre, next to the core, {\it before the
  mean temperature of the fragment rises to} $\sim 2000$~K .

This instability is quite analogous to that laying the foundation of the core
accretion plus gas capture theory of giant planet formation
\citep[e.g.,][]{PollackEtal96} , which we refer to as simply core accretion
(CA) hereafter. In CA, the core grows by accretion of planetesimals from the
disc and gravitationally attracts a gaseous "atmosphere", again from the
disc. The atmosphere's mass increases as a steep power of core mass
\citep{Stevenson82}. When the two masses become comparable, there is no stable
hydrostatic solution and the atmosphere collapses onto the core dynamically
\citep{Mizuno80}, triggering a rapid gas accretion from the disc onto the
protoplanet \citep[see][for a recent detailed treatment of the
  problem]{Rafikov06}. We find a very similar sequence of events, except that
the core grows by accretion of small grains rather than planetesimals, and all
the action occurs inside a self-gravitating gas fragment born from GIs.

Understanding the core accretion-like instability discovered here is a key to
understanding the pre-collapse evolution of the disc instability planets, and
is necessary for identifying ways to distinguish observationally between
planet formation modes. To differentiate between the instability discussed
here and the one in the CA hypothesis, we call the collapse studied in this
work a "core-assisted gas capture" (CAGC) instability, with emphasis that the
core is embedded deep within a massive self-gravitating gaseous
clump\footnote{Hereafter, we refer to the gaseous region immediately
  surrounding a sedimented core as the core's ``atmosphere'', even though it
  is deep within a fragment's interior.}.

Global disc fragmentation simulations are not yet advanced enough to follow
both the formation of a clump and the clump's subsequent evolution to
planetary scales.  In particular, the region near the growing core is
numerically challenging to investigate due to the required resolution and the
extreme changes in gas composition.  For example, a core will have a radius of
$\sim 10^{-4}$~AU, which is about 10,000 times smaller than the size scale of
a molecular hydrogen gaseous clump.  Although this core may be small compared
with the total extent of the clump, we shall later see that destabilisation of
gas directly surrounding the core can lead to collapse of the entire system.

Here, as a first step, we present a series of simple 1D models that focus on
the structure of the gas atmospheres near cores that have a range of core
properties (masses and luminosities). Our primary purpose is to address
the following: (a) the conditions necessary for a core within a
self-gravitating fragment to prompt dynamical collapse of the core's
atmosphere; (b) the critical core's mass at which this collapse happens, and
(c) the likelihood of these conditions to be met by fragments born from GIs
within a reasonable protoplanetary disc setting. We present our numerical
methods in section 2, and show the results of a case study for a particular
clump mass in section 3.  We then explore in section 4 the parameter space for
which cores can drive the collapse of their atmospheres.  Our results are
summarised and their implications are discussed in section 5.

\section{Physics and numerical method}\label{sec:method}

\subsection{Relation to previous literature}\label{sec:motivation}

The basic physics of core-assisted gas capture instability is very similar to
that of the core accretion instability \citep[e.g.,][]{Mizuno80}. Consider a
very dense core (solid or liquid) immersed in an initially uniform gas of a
given (lower) density and temperature. Gravitational attraction of the core
will pull gas layers closest to the core even closer, compressing the
material, until a hydrostatic balance is established. At the same time,
compression heats the material up, and since this all occurs on the (very
short) local dynamical time scale, the compressed layer initially has an
adiabatic structure. \cite{PerriCameron74} questioned hydrodynamical stability
of the compressed gas, showing that when the core mass exceeds a critical
value, $M_{\rm crit}$, the layer becomes unstable due to combined core and
self-gravity and will collapse to much higher densities. The critical core
mass was found to be as high as $\sim 100 \mearth$ for conditions thought to
be typical in the Solar Nebula \citep[see also][]{Wuchterl93}.

However, two complications arise: (i) radiation diffusion may transport the
compressional heat out, presumably decreasing the temperature of the
compressed gas and hence $M_{\rm crit}$ too; (ii) On the other hand, the core
may be a significant additional source of heat that acts in the opposite
direction. A number of authors
\citep{Harris78,MizunoEtal78,Mizuno80,Stevenson82} subsequently relaxed the
assumption of adibaticity for the gas in the envelope around the
core. Assuming thermal energy equilibrium in the envelope instead, they
obtained the critical core masses spanning a broader range from $\sim 1
\mearth$ to $\sim 100 \mearth$, depending on the outer boundary condition and
opacity in the envelope\footnote{The absolute lower limit to the critical core
  mass is obtained \citep[e.g.,][]{Sasaki89,Nayakshin10b} by assuming an
  isothermal structure for the envelope, $M_{\rm iso} \approx 1 \mearth \;
  T_3^{3/2} (\mu/2.45 m_p )^{-3/2}$ where $T_3= T_{\infty}/10^3$~K is the
  envelope's temperature in units of 1000~K and $\mu$ is the mean molecular
  weight of the gas. This result is only pedagogically interesting, however,
  as the isothermal assumption is unrealistic for most conditions explored
  here: the cooling time within the dense envelope near the core can exceed
  the cooling time of the protoplanet itself and even the age of the
  Universe. Real systems will therefore never reach the isothermal solution
  and require larger $M_{\rm crit}$.}. See \cite{Rafikov06} and
\cite{HoriIkoma11} for more recent studies of the problem.

These previous studies inform us here and shape our approach. A key
  difference between this work and what has been done previously is that we
  focus on massive clumps born by disc instability.  Furthermore, we assume
  that dense cores readily form in these clumps as a result of grain settling.
  Strictly, the core formation timescale depends on the grain size
  distribution, but is estimated to be of the order of a few thousand to a few
  $\times 10^4$ years (Helled et al., 2008; Nayakshin, 2010), which we use in
  our calculations below. Within this context, we explore the hydrostatic
  stability of gas near the interface between the dense core and the gaseous
  envelope.  Finally, for the calculations here, we do not simultaneously
  investigate the full evolution of fragments from their formation.  Instead,
we isolate the innermost region within evolved clumps (fragments) for which an
embedded core has an immediate influence.  We define this radius of influence
as $r_i = GM_{\rm c}/c_\infty^2$, where $c_\infty$ is the gas sound speed at
``infinity'', which is the bulk of the mass of a fragment.  The region within
$r_i$ defines the atmosphere of the core, which is typically $r_i \sim
10^{-2}$~AU for core masses $\sim 10$ M$_{\oplus}$.

The presence of the envelope mass at $r>r_i$ is treated as an outer boundary
condition. We assume that the gas fragment itself is not tidally limited.
Because initial clump sizes can have radii over an AU in extent, our
assumption of negligible tides implicitly assumes that the clumps are
relatively far from their stars. Our calculations are relevant as long as the
fragment is dominated by molecular hydrogen and does not fill its Roche lobe,
which implies generally a distance to the host star of at least $1$~AU or
more, depending on the mass and age of the fragment \citep[see,
  e.g.,][]{Nayakshin10c}. As an example, for a $10^4$ years old fragment of 2
M$_J$, the corresponding orbital distance must be larger than $\sim$ 10 AU. 

In a fragment with radius $R \sim O(1 \hbox{AU})$, the gas density at $r_i \sim
0.01 R$ is equal to the central clump density and is higher than the mean
density. Tidal forces are thus negligible inside $r_i$. This is in contrast to
the CA instability inside a protoplanetary disc, where the outer boundary of
the gas atmosphere around the core can be set by the Hill's radius.

\subsection{Computational procedure}\label{sec:overall}

For simplicity, we assume a constant density for the core's internal structure
with $\rho_{\rm c} = 10$~g~cm$^{-3}$, which reflects the expected density
  of giant planet cores \citep[e.g.,][]{Guillot05}. The sensitivity of the
  results to the chosen core density is investigated below, and found quite
  weak.  The corresponding core radius is given by $r_{\rm c} = (3 M_{\rm
  c}/4\pi \rho_{\rm c})^{1/3}$.  Because a fragment is not directly evolved in
these models, and to reduce parameter space of our first study, we use the
approximate formulae from \cite{Nayakshin10a,Nayakshin10c} to describe gas
density and temperature profiles at any given time $t$ after the formation of
a fragment, which can be used to derive initial conditions and boundary
values.  Eventually, we will incorporate the models presented here with
simulations that also take into account the global evolution of clumps.

Our next steps are identical to what is done in CA studies except for the
assumed core luminosity and atmosphere's metallicity structure, as explained
in \S \ref{sec:accretion} below.  Starting with the boundary conditions from
our evolved fragment models (actual evolution times given below), we integrate
the equations of hydrostatic and energy transfer equilibrium inward from
$r=r_i$ all the way to the core radius, $r=r_{\rm c}$, using the standard
procedure for atmospheric solution finding \cite[e.g., see][for a recent
  treatment of the problem]{HoriIkoma11}.  In particular, the hydrostatic
balance equation reads
\begin{equation}
{dP \over dr} = - {G \left( M_{\rm c} + M_{\rm atm}(r)\right)\over r^2}\rho\;,
\label{hstatic}
\end{equation}
where $M_{\rm c}$ is the mass of the core, $P$ and $\rho$ are the gas,
pressure, and density, respectively, and $M_{\rm atm}(r)$ is the total mass of
gas inside radius $r$.  The gas includes an admixture of heavy elements.

Energy dissipation within the atmosphere itself is not included in these
  calculations.  Although gravitational contraction will lead to a significant
  increase in a clump's thermal energy over time, the energy released into the
  clump from gravitational work is deposited throughout the system, with only
  a small fraction deposited near the core. Since collapse of the material
  immediately surrounding the core (its atmosphere) occurs when its mass is of
  order the core's mass and the atmosphere's radial extent is a few to ten
  times that of the core (cf. figures 1 and 2 below), the total binding energy
  reservoir of this region must be considerably smaller than the binding
  energy of the core. This implies that converting the atmosphere's binding
  energy into compressional heat is unlikely to result in luminosity exceeding
  that of the core, except perhaps for a short time. As a result, the
  luminosity that passes through the atmosphere is, to a good approximation,
  set by the luminosity of the core itself:
\begin{equation}
L_c = \hbox{ const}\; = -{64 \pi r^2 \sigma_B T^3 \over 3\kappa\rho} {d T
  \over dr} \;,
\label{Latm}
\end{equation}
where $\rho$ and $T$ are the gas density and temperature at a distance $r$
away from the core's centre. Opacity coefficient $\kappa$ is a function of gas
density and temperature, and is taken from \cite{ZhuEtal09}, which is an
updated version of \cite{Bell94} opacities. The treatment of $L_c$ will be
discussed below.  The energy transport mechanism in the atmosphere assumed in
equation (\ref{Latm}) is radiative diffusion.  However, if the actual
temperature gradient, $\nabla_{\rm rad} = (P/T) dT/dP$, as calculated from
equations (\ref{hstatic}) and (\ref{Latm}), is larger than the adiabatic
gradient (the convective stability criterion), $\nabla_{\rm conv}$, the
temperature gradient is limited to $dT/dP = (T/P) \nabla_{\rm conv}$.

The adiabatic temperature gradient is computed directly from the EOS of the
mixture.  The EOS we use was kindly provided by A.Kovetz and is similar to the
one recently presented by \cite{VazanEtal13}.  For hydrogen and helium the
SCVH \citep{SaumonEtal95} EOS is used, while for lower pressures the EOS was
extended based on the Debye approximation for a weakly interacting
mixture. For the heavy elements, which are here represented by SiO$_2$, the
EOS is based on the quotidian equation of state (QEOS) described in
\cite{MoreEtal98}. The density of the mixture is calculated by applying the
additive volume law, where the density of the mixture is determined by
assuming that the volume of the mixture is the sum of the volumes of the
individual components. The entropy of the mixture is the sum of the entropies
of the individual components plus a term for the entropy of mixing
\citep{SaumonEtal95}.

The equations of hydrostatic and energy balance just described allow us to
integrate the density and temperature profile inwards all the way to the
core's boundary, provided that $M_{\rm atm}(r)$ is known. This is not actually
the case, and therefore iterations are required. Setting $M_{\rm atm}(r)=0$,
we integrate the above equations from $r_i$ to $r_c$ once, and then repeat the
procedure with $M_{\rm atm}(r)$ calculated in the previous iteration. This
procedure either converges to some finite value of the atmosphere's
  mass, yielding a stable structure inside $r_i$ -- an atmosphere bounded by
a massive envelope and a central core, or diverges, resulting in a runaway in
the gas density and in the enclosed mass in the atmosphere.  The largest value
of $M_{\rm c}$ at which a stable atmospheric solution exists represents the
critical core mass, $M_{\rm crit}$.

\subsection{Metal pollution and the critical core mass}\label{sec:accretion}

Cores can grow at the centre of clumps through the sedimentation of small
grains, e.g., $\sim$ a few cm.  In our previous work
\citep[e.g.,][]{HS08,Nayakshin10a,Nayakshin10b}, these grains were assumed to
accrete onto a core if they entered the innermost Lagrangian gas mass zone,
below which the core is located. However, resolving the gas temperature
profile closer to the core may invalidate this assumption since gas near the
core is hotter.  Grains evaporate rapidly when the surrounding temperature
exceeds the grain's vaporisation temperature, $T_{\rm v}$, for the given
material.  Even for most of the refractory abundant grains, such as rocks,
$T_{\rm v} \approx 1400$~K \citep[e.g.,][]{PodolakEtal88}, with the rate of
evaporation depending on grain size.  As we show below, the temperature of the
gas near the core is higher than 1400 K for all relevant parameter space
considered, e.g., when the core mass increases above $\sim 1 \mearth$. We note
in passing that this also implies that the dense cores of heavy elements that
we study here are not necessarily solid.  We simply assume that the core
consists of heavy elements and that its density exceeds the surrounding gas
density by orders of magnitude, and this is what matters for, e.g., setting
the inner boundary of the atmosphere, $r_{\rm c}$. Whether the core is solid,
liquid, vapour, or (the most likely) a mix of these three phases is however
important in determining the luminosity of the core, as discussed below.

When the core grows sufficiently massive, grains do not reach the core, but
instead sediment down to the innermost region where $T > T_{\rm v}$ and
evaporate there.  The heavy element content and hence the mean molecular
weight of the gas around the core will increase above that of more distant
regions (where $T < T_{\rm v}$ and heavy elements continue to sediment as
grains).  We shall also see that the hot, inner, heavy-element-rich layer is
usually strongly dominated by convection.  Due to convective mixing, the heavy
element material in this region is well mixed with the gas throughout the
convective layer, homogenising the heavy element content
\citep[cf. also][]{HoriIkoma11}.

\cite{HoriIkoma11} have used a similar setup to calculate the critical core
mass in the context of the CA model for planet formation.  It was found that
the layer ``polluted'' by grains, formed in a way similar as described above,
significantly reduces $M_{\rm crit}$ in the limit that the gas becomes
dominated by heavy elements.  In particular, $M_{\rm crit}$ may even fall
below $1\mearth$ for heavy element fractions of $z\simgt 0.8$ of the total gas
mass.  The main physical reason for such a behaviour is the increase in the
mean molecular weight of the gas-heavy element mixture at high $z$ (although
latent heat, such as that due to evaporation of grains also helps to reduce
$M_{\rm crit}$).

A similar result is found here.  However, we note that at very high-$z$,
collapse of a relatively low-mass ($M\simlt 1\mearth$) heavy-element-rich
layer onto the core does not, in general, mark the collapse of the entire
fragment.  The gas above the metal-polluted layer has metallicity much closer
to the initial heavy element abundance of the gas, although depleted in
accordance with the enrichment of the inner layers.  The critical core mass
for the gas with a Solar-like heavy element abundance is much higher.
Therefore, collapse of the polluted heavy-element-rich atmosphere onto the
core should strictly be considered as accretion of the inner layer onto the
core.  The accreting material is dominated by solids in this regime (because
$z\rightarrow 1$), and only a small fraction $(1-z)$ of it is hydrogen and
helium.  The collapse of the metal-polluted layer is thus effectively a
delayed accretion of metals onto the core, which is not very different from a
direct accretion of grains onto the core.  Similar conclusions were reached by
a number of authors in the context of CA model, although incoming high-z
material there is in the form of planetesimals \citep[e.g.,][]{Mordasini13}.
On the other hand, the luminosity of the core may be significantly different
in these two pictures, as further discussed in section
\ref{sec:core_luminosity}.

We thus arrive at the following phases of the core growth: (1) When the mass
and the luminosity of the core is low, grain material is able to impact the
core directly, as assumed in some previous studies.  (2) When the core mass
increases, eventually the gas near the core is hot enough to evaporate the
grains.  As a result, grains can evaporate before reaching the core.  Their
destruction simultaneously creates a heavy-element-polluted layer.  As time
progresses, the layer becomes completely dominated by high-z materials and can
thus collapse onto the core, increasing its mass.  Note that some H/He also
accretes onto the core in this process.  (3) As the core grows in mass, less
heavy element pollution is required for the high-z layer to collapse.
Eventually, layers with $z\sim 0.5$ are liable to collapse (this will be
calculated below).  At this point the core starts to gain as much mass in H/He
as in metals, and this marks the beginning of rapid gas accretion onto the
core.

Therefore, we define the critical core mass as one at which a moderately
metal-polluted layer, $z=0.5$, can collapse onto the core.  After such a layer
collapses, the increased mass of the core enables accretion of relatively
metal-poor gas, $z < 0.5$, for which the presence of metals makes only a minor
difference.

\subsection{Core Luminosity}\label{sec:core_luminosity}

 In this section we discuss the luminosity of the core, which can play a
 significant role in setting $M_{\rm crit}$.  In the core accretion model, it
 is customary to write
\begin{equation}
L_{\rm core} = {G M_{\rm c} \dot M_{\rm c}\over r_{\rm c}}\;,
\label{l_CA}
\end{equation}
where $\dot M_{\rm c}$ is largely the planetesimal accretion rate onto a core.
As discussed in the previous section, small grain accretion is necessarily
complicated by grain evaporation and mixing with H/He in the
heavy-element-polluted layer, which is also true in core accretion.
Furthermore, when the heavy-element-rich layer does collapse onto the core, it
is unclear how quickly the compressional heat gained in the collapse will
escape from the core region.

These uncertainties are physically related to the difficult issue of
determining how much heat is retained by gas that is accreted onto low-mass
stars \citep[e.g.,][]{PrialnikLivio85,HartmannEtal11} or onto giant planets
\citep[e.g.,][]{MarleyEtal07}.  One can parameterise the energy retained per
unit mass of accreted matter as $\alpha G M_{\rm c}/r_{\rm c}$, where $\alpha$
is an unknown factor between 0 and 1.  Different values of $\alpha$ can lead
to very different evolution histories, e.g., for stellar radius and luminosity
in the two opposite limits \citep[e.g.,][]{PrialnikLivio85}.  When $\alpha \ll
1$, the accretion is called ``cold", and models that use these conditions have
a "cold start."  When $\alpha \simlt 0.5$ (as opposed to being very small,
i.e., $\alpha \ll 1$), the accretion is ``hot'', and corresponding models have
a "hot start."

While there are many similarities between core accretion and core
sedimentation, there are also important differences.  When core formation
occurs within a gaseous clump as studied here, the core is buried deep in a
very optically thick gas.  As a result, the energy cannot be radiated away
easily.  To reflect this change, and the uncertain energy release rate under
the accretion of heavy-element-polluted layers, we parameterise the energy
loss from the core as
\begin{equation}
L_{\rm core} = {G M_{\rm c}^2 \over r_{\rm c} t_{\rm kh}}\;,
\label{l_kh}
\end{equation}
where $t_{\rm kh}$ is the Kelvin-Helmholtz contraction time of the solid core.
The value of $t_{\rm kh}$, which is taken to be a free parameter in our model,
is varied over a broad range below.

Although the exact value of $t_{\rm kh}$ is unclear, there is a lower limit on
$t_{\rm kh}$ imposed by the assembly time of the core.  For example, compare
equation (\ref{l_kh}) with equation (\ref{l_CA}), and let us also assume that
all the accreted heat is being rapidly radiated away from the core.  In this
case $t_{\rm kh}$ should be replaced by the time scale on which the core
grows, $t_{\rm growth} = M_{\rm c}/\dot M_{\rm c}$, and the accretion
luminosity given by equation (\ref{l_CA}) is recovered.  From evolutionary
calculations of fragments \citep[e.g.,][]{HS08,Nayakshin10a}, $t_{\rm growth}$
is from a few $\times 10^3$ to $\simlt 10^5$ yr.  For this reason, we take
$10^4$ yr as a rough representative cooling timescale.  We do note that
understanding the cooling timescale of clumps is an active area of research,
with some groups finding, for certain conditions, very fast cooling times \citep{HB11}.

On the other hand, the heat may be trapped inside the core for times much
longer than $t_{\rm growth}$.  To explore a low-luminosity limit, we also
consider values of $t_{\rm kh}$ in the range of $10^7$ yrs.

\section{An Example - 2 M$_J$ Fragment}\label{sec:examples}

We now consider an isolated gas clump of mass $M = 2 M_J$ at time $t_{\rm age}
= 10^4$ yr from the fragment's birth. This is motivated by the typical cooling
time scale for the clumps, and is also consistent within an order of magnitude
with the clump migration timescale into the inner disc
\citep[e.g.,][]{Nayakshin10c}. The fragment is dominated by molecular
hydrogen, so that the mean molecular weight at $r \ge r_i$ is set to $\mu =
2.45 m_p$.  The typical clump densities and temperatures in the centre of our
clumps are estimated by using the analytical study of \cite{Nayakshin10a}.  In
particular, we set the clump radius $R = 0.8$~AU~$(t_{\rm age}/10^4)^{-1/2}$.
The temperature and mean density of the clump are then $GM\mu/(2 k_B R)$ and
$3 M/ (4\pi R^3)$, respectively.  This yields gas density $\rho_\infty \approx
5.5\times 10^{-10}$~g~cm$^{-3}$ and temperature $T_\infty \approx 330$~K as
the outer boundary conditions for our core atmosphere calculation.  We plan to
present more detailed time-dependent 1D models with different grain species in
the near future.

\begin{figure}
\psfig{file=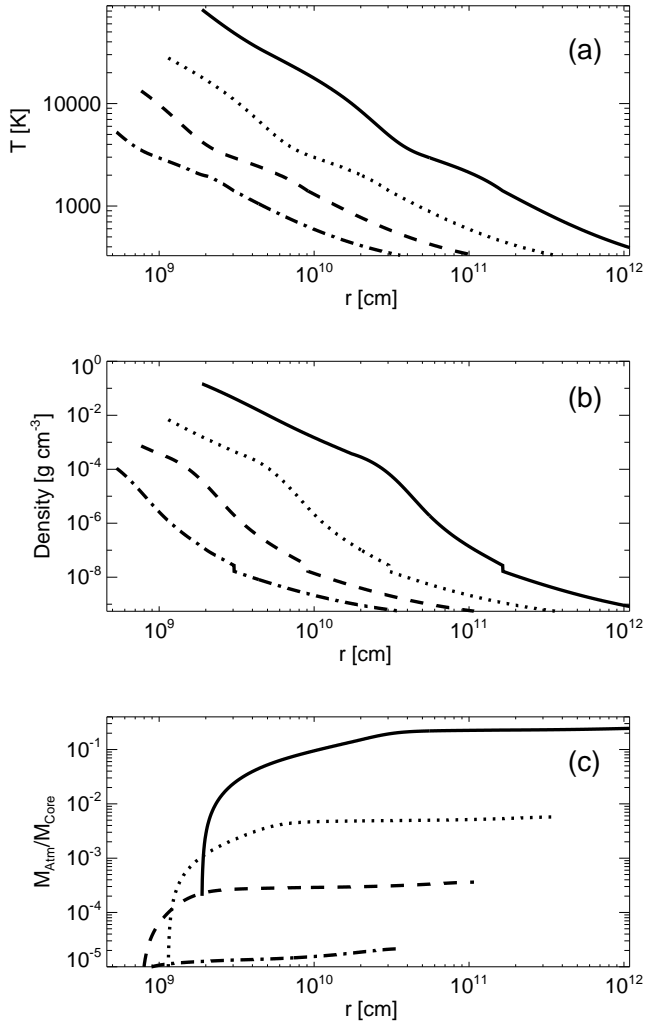,width=0.5\textwidth,angle=0}
\caption{Structure of the atmosphere around cores of different masses in the
  bright core limit ($t_{\rm kh}=10^4$ years). The curves show gas temperature
  (panel a), gas density (b), and the enclosed mass $M_{\rm atm}(r)/M_{\rm c}$
  (c). The core mass is $M_{\rm c}$ = 1, 3, 10 and 45~$\mearth$ for the
  dash-dotted, dashed, dotted and solid curves, respectively. The
  discontinuity in density marks the division between the inner metal rich
  $z=0.5$ atmosphere and the outer "normal" metal abundance gas, $z=0.02$.
  There is no steady-state atmospheric solution for $M_{\rm c}$ larger than
  45~$\mearth$; the atmosphere collapses hydrodynamically.}
\label{fig:fig1}
\end{figure}

Figure \ref{fig:fig1} presents the atmosphere structure for four different
core masses, $M_{\rm c} = 1$, 3, 10 and 45~$\mearth$, shown with dot-dashed,
dashed, dotted, and solid curves, respectively.  All of these are computed in
the limit of a relatively short $t_{\rm kh}$, which we set to $t_{\rm kh} =
10^4$~yrs.  Panels (a) and (b) show the gas temperature and density structure,
respectively.  As expected, the atmosphere near the solid core becomes denser
and hotter as $M_{\rm c}$ increases.  Notice that the radial range of the
atmosphere also increases.  The inner boundary of the atmosphere, $r_{\rm c}$,
is proportional to $M_{\rm core}^{1/3}$, resulting from the assumed fixed core
density.  The outer boundary $r_i\propto M_{\rm core}$.  Note the jump in
density by about a factor of 2 in panel (b).  This is the point where the
metallicity of the gas is assumed to jump discontinuously from $z = 0.02$ ($T
< T_{\rm v}$, outer regions) to $z=0.5$ ($T > T_{\rm v}$), inner region.  The
gas temperature and pressure, $P = (\rho/\mu) k_B T$, are continuous across
this transition, but the gas density $\rho$ and the mean molecular weight
$\mu$ are both discontinuous by the same factor across the kink.

Panel (c) of Figure \ref{fig:fig1} displays the ratio of the enclosed mass of
the atmosphere, $M_{\rm atm}(r)$, to the core mass, $M_{\rm c}$.  This ratio
is found to increase strongly with $M_{\rm c}$.  The three lower core mass
cases in Figure \ref{fig:fig1} show stable atmospheres that increase in mass
proportionally with the core mass.  The fourth case, which is the highest core
mass case (solid curve), however, is qualitatively different.  The atmosphere
mass in this case is comparable to that of the core.  At a core mass just
slightly larger than 45~$\mearth$, we find no static atmosphere solution; the
atmosphere collapses dynamically.  Therefore, the critical core mass in the
$M=2$ M$_J$, short $t_{\rm kh}$, case is $M_{\rm crit} =45 \mearth$.

\begin{figure}
\psfig{file=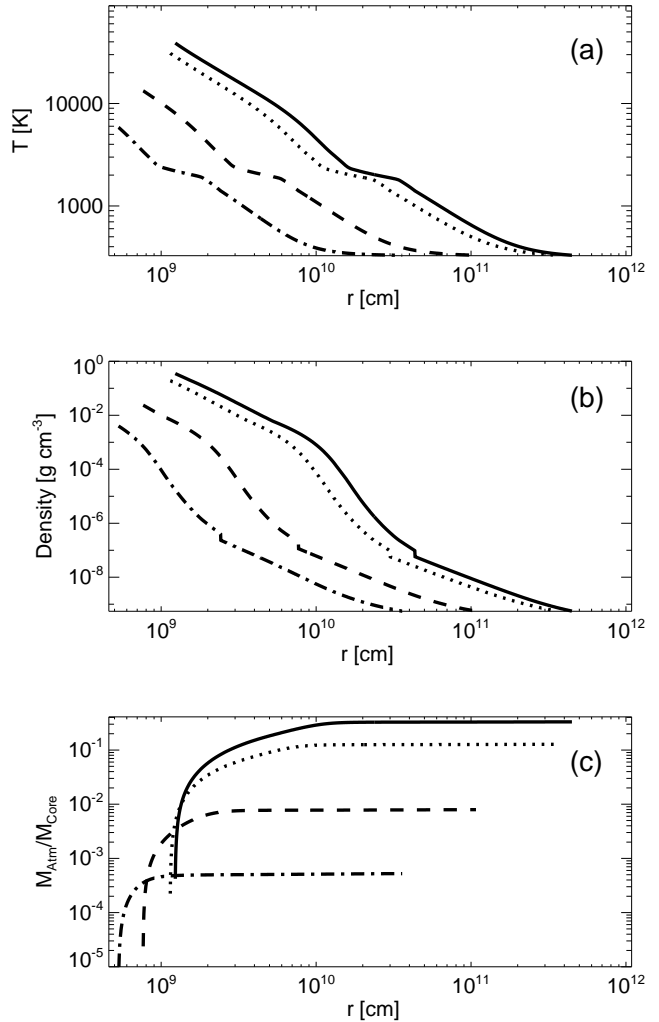,width=0.5\textwidth,angle=0}
\caption{Same as Fig. \ref{fig:fig1}, except in the dim core limit ($t_{\rm
    kh}=10^7$ years). The core mass is $M_{\rm c}$ = 1, 3, 10 and
  12.5~$\mearth$ for the dash-dotted, dashed, dotted and solid curves,
  respectively. There is no steady-state atmospheric solution for $M_{\rm c}$
  larger than 12.5~$\mearth$; the atmosphere collapses hydrodynamically.}
\label{fig:fig2}
\end{figure}

Figure \ref{fig:fig2} shows a very similar calculation but for a long $t_{\rm
  kh}$ case, $t_{\rm kh} = 10^7$~yr.  The core masses in this case are $M_{\rm
  c}$ = 1, 3, 10 and 12.5~$\mearth$.  By comparing Figures (\ref{fig:fig2})
and (\ref{fig:fig1}), it is found that the atmosphere is much more dense in
the long $t_{\rm kh}$ case models, as expected: brighter cores require hotter
atmospheres to transport the heat flux away.  For this reason, the atmospheric
collapse occurs at $M_{\rm c}\approx 12.5\mearth$, which is at a much lower
core mass than in the short $t_{\rm kh}$ case.

To put these values of $M_{\rm c}$ in perspective, we note that a gas clump of
mass $2 M_J$ with solar [Fe/H] will have $\approx 12 \mearth$ of heavy
elements. In order to get a core of $45 \mearth$ we need a gas clump
metallicity which is more than 3 times solar, or the clump needs to be
  significantly enriched relative to its birth nebula through, for example,
  planetesimal capture or gas-solid aerodynamic effects 
\citep[e.g.,][]{BoleyEtal11a}.

\section{Parameter space of collapsing atmosphere models}\label{sec:param}

We have illustrated the computational procedure with which we can determine
the properties of the atmosphere of a heavy element core grown inside a
fragment for a fixed set of properties, e.g., $\rho_\infty$ and $T_\infty$.
This procedure could be used in a fully self-consistent calculation for which
the properties of the gas clump and the rate of grain sedimentation are
calculated as a function of time.  The critical core mass then depends on the
time evolution of the clump, and the value needed to induce collapse may or
may not be reached.  For example, the gas clump may contract and heat up too
quickly for significant grain sedimentation and core formation to take
place \citep{HS08,Nayakshin10a}.

In a yet more self-consistent calculation, the clump properties depend on its
changing environment as it migrates inward.  As noted in section
{\ref{intro}}, the clump may be even destroyed by tidal forces from the parent
star \citep[e.g.,][]{BoleyEtal10,Nayakshin10c} or by over-heating due to the
thermal bath effect from the surrounding disc and stellar irradiation
\citep[e.g.,][]{CameronEtal82,VazanHelled12} well before a core ever forms.

We have already implemented some of these model developments.  However, the
complexity and the parameter space of such calculations are very large, and
they fully deserve a separate future study.  In this section we thus present a
more modest approach in which we study isolated fragments that contract with
time as prescribed by the simple analytical model of \cite{Nayakshin10c}.
This should provide a glimpse on how the critical core mass for atmosphere
collapse scale with properties of the gas clumps, e.g., their masses.  Such a
mapping, together with the simple budget constrains on solids in a given
clump, may allow us to make some {\it preliminary} conclusions as to when
conditions needed for the atmosphere collapse can be realistically reached.

\begin{figure}
\psfig{file=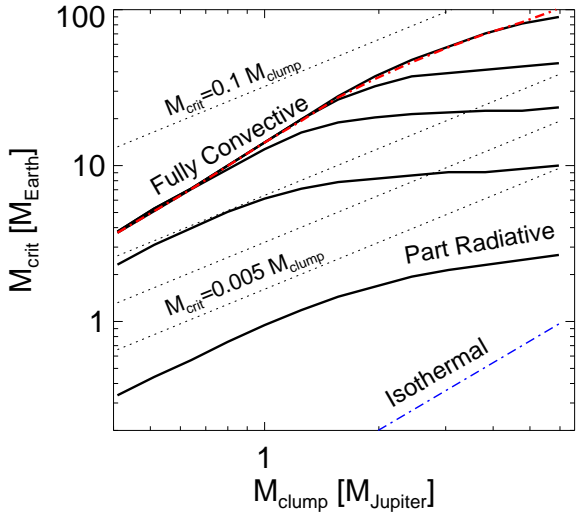,width=0.5\textwidth,angle=0}
\caption{The critical core mass, $M_{\rm crit}$ versus the gas mass of the
  clump, $M_{\rm cl}$, shown with thick curves.  The five black solid curves
  are for different values of the core cooling time, $t_{\rm kh} = 10^4, 10^5,
  10^6, 10^7$ and $10^8$~yr, from top to bottom, respectively. The thick
  dot-dashed (red) curve is same but for $t_{\rm kh} = 10^3$~yrs.  The
  dot-dashed (blue) line on the bottom shows the ``isothermal limit'', $L=0$
  (see the footnote in \S \ref{sec:motivation}).  The power law dotted lines
  show constant $M_{\rm crit}/M_{\rm clump} = $0.005, 0.01, 0.02, and 0.1,
  from bottom to top, respectively.  These lines reflect the minimum heavy
  element abundance required to initiate a CAGC instability.  The red
  dot-dashed curve shows the limit at which the atmosphere becomes fully
  convective, which is why the curves become more closely spaced for higher
  core luminosities (small $t_{\rm kh}$).}
\label{fig:param1}
\end{figure}

We use the same simple model, fixing the age of the gas clump at $t_{\rm age}=
10^4$~yr as before, but we now consider a range of gas clump masses.  Gas
clumps much less massive than $\sim 1 M_J$ are difficult to form in
protoplanetary discs through fragmentation \citep[e.g.,][]{BoleyEtal10}, while
clumps more massive than $\sim 6 M_J$ may be too hot at $t\sim 10^4$~yrs to
allow grain sedimentation ($T_\infty > 1400$ K).  For these reasons, we
constrain our parameter search to be between 0.4 and 6 M$_J$.  Figure
\ref{fig:param1} shows the critical core mass as a function of the total gas
mass of the clump, $M_{\rm clump}$, for several different values of the core
cooling time, $t_{\rm kh} = 10^4, 10^5, 10^6, 10^7$ and $10^8$~yr, depicted by
the solid curves from top to bottom, respectively.  The highest luminosity
case considered, $t_{\rm kh}=10^3$~yr is shown with a dot-dash red curve to
distinguish it from the $t_{\rm kh} = 10^4$~yr case.

The thin dash-dotted (blue) line on the bottom of the figure, marked
"Isothermal", shows the isothermal critical core mass, $M_{\rm iso}$,
corresponding to the non-accreting and non-radiating core
\citep[e.g.][]{Nayakshin10b}. This $L=0$ limiting case is non-realistic
  as gas cooling time near the surface of the core may be too long for the
  thermal equilibrium to be achieved (see the footnote in \S
  \ref{sec:motivation}), but it is nonetheless instructive as an (analytical)
  absolute lower limit to $M_{\rm crit}$. This line marks the estimated
minimum critical core mass for a given temperature and density in the fragment
($T_\infty$ and $\rho_\infty$). $M_{\rm iso} \propto T_\infty^{3/2}$, with a weaker
dependence on $\rho_\infty$, this translates into a roughly $M_{\rm iso}
\propto M_{\rm clump}^{3/2}$ scaling for the isothermal limit on the figure,
given the simple \cite{Nayakshin10c} model for the gas fragment.

Figure \ref{fig:param1} shows, as expected, that any $L>0$ core ($t_{\rm
    kh} < \infty$) requires a core mass larger than $M_{\rm iso}$ to enforce
atmosphere collapse because the atmosphere around a luminous core is
hotter. The least bright core case studied here, $t_{\rm kh} = 10^8$ yrs, is
the solid curve that gives the lowest values of $M_{\rm crit}$.  Analysis of
atmospheric profiles of the respective models shows that the atmospheres on
the flat part of the solid curves in fig. \ref{fig:param1} (marked by "Part
Radiative" in the figure) contain both convective and radiative regions.  We
should expect in this regime that the brighter the core is (the shorter
$t_{\rm kh}$), the more massive the core must be to cause the atmosphere to
collapse.  This trend is to be expected from the well known radiative zero
solution for atmospheres \citep[e.g.,][]{Stevenson82}.  The atmosphere mass
scales as $M_{\rm atm}\propto M_{\rm core}^4/L$, so that higher $L$ reduces
$M_{\rm atm}$ at a given $M_{\rm core}$.  This explains why the critical core
mass increases with decreasing $t_{\rm kh}$ on the flat part of the solid
curves. The reason that the curves in this regime are almost independent of
the clump mass, $M_{\rm clump}$, is that the radiative zero solutions are
dominated by regions very close to the cores, and are hence insensitive to
$\rho_\infty$ and $T_\infty$.

Sensitivity of $M_{\rm crit}$ to the core luminosity however becomes weaker
and weaker as $t_{\rm kh}$ decreases, particularly for low-mass clumps.  This
is obvious from the top few solid curves in the figure, which become very
closely spaced.  In fact, models with $t_{\rm kh}\le 10^4$~yr saturate at a
single curve (dash-dot red).  The atmosphere of this limiting solution is
fully convective, so that $dT/dr$ is independent of $L$ everywhere. The
dash-dot red curve also marks this fully convective limiting solution. 
  The solution corresponds to the fully convective, and thus adiabatic,
  atmospheres studied first by \cite{PerriCameron74}, and also leads to very
  large values for $M_{\rm crit}$. Considering the upper curves in figure
\ref{fig:param1}, we find that the approximate relation $M_{\rm crit} \approx
0.045 M_{\rm clump}$ matches the fully convective limit within $\sim 30$\%.
The linear behaviour of $M_{\rm crit}$ with increasing $M_{\rm clump}$ can be
obtained analytically by assuming that the atmosphere remains adiabatic with a
constant specific heats ratio, $\gamma$ (which is not exactly the case here,
explaining why the curve is not quite linear).

The dotted thin lines in figure (\ref{fig:param1}) are lines of a constant
ratio $M_{\rm crit}/M_{\rm clump} = $0.005, 0.01, 0.02, and 0.1, from bottom
to top, respectively. These lines are useful in determining the parameters of
the clumps that may host a core that is massive enough to initiate a
core-assisted gas capture (CAGC) instability at a given clump metallicity.  If
the curve (of a fixed $t_{\rm kh}$) is above the line, then it is not possible
to trigger CAGC for the given heavy element fraction and total clump mass,
even if all the heavy elements sedimented into the core.  For example, at half
of the solar heavy element abundance (second from bottom dotted line), the
maximum core mass would be $\sim 0.01 M_{\rm clump}$.  Under these conditions,
only gas clumps more massive than $\sim 3 M_J$ could possibly provoke collapse
and only if $t_{\rm kh} = 10^7$~yr (second from bottom solid curve) or
longer.  If $t_{\rm kh}\le 10^6$~yr then no clump of this metallicity could
ever collapse via the CAGC instability.

Evidently, assembling a core that is massive enough to induce atmosphere
collapse is very difficult for systems that have low nominal heavy element
fractions (i.e., low metallicity systems).  Increasing the available budget of
grains clearly improves the chances of reaching the critical core mass,
creating a potentially strong dependence on heavy element abundance.
Nonetheless, the exact form of this dependence is unclear from our study
alone, and requires future additional work.

\section{Sensitivity to assumptions}\label{sec:sensitivity}

\subsection{Core density}

\begin{figure}
\psfig{file=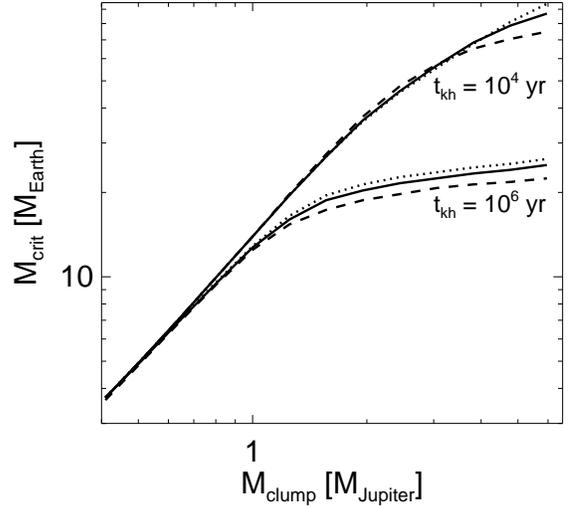,width=0.5\textwidth,angle=0}
\caption{Sensitivity of the critical core mass, $M_{\rm crit}$, to the core
  density. The solid curves are the same as the respective curves in Figure
  \ref{fig:param1} ($t_{\rm kh} = 10^4$ and $10^6$ yr, as labelled on the
  figure). Recall that these two curves were computed assuming core density
  $\rho_{\rm c} = 10$~g~cm$^{-3}$. The dotted and the dashed curves are for
  $\rho_{\rm c} = 30$~g~cm$^{-3}$ and $\rho_{\rm c} = 1$~g~cm$^{-3}$,
  respectively, with all other parameters the same as in the solid curves}
\label{fig:rho_core}
\end{figure}

In this paper we have assumed a fixed core density, $\rho_{\rm
  c}=10$~g~cm$^{-3}$. Density of the core could in  principle be higher than
this for very massive cores due to their large self-gravity. Alternatively,
the core density could also be significantly lower than assumed here, especially
for young hot cores (say, $t_{\rm kh}\simlt 10^4$ yr) which have not yet
been able to radiate away the primordial heat of their formation. To test
sensitivity of our results to the assumed value of $\rho_{\rm c}$, we rerun
some of the calculations presented in figure \ref{fig:param1} for two
additional values of core density, $\rho_{\rm c} = 30$~g~cm$^{-3}$, and
$\rho_{\rm c} = 1$~g~cm$^{-3}$. Figure \ref{fig:rho_core} shows the results of
this experiment with the dotted and the dashed curves, respectively, for these
two values of the core densities. The solid curves reproduce the corresponding
curves from figure \ref{fig:param1}.

We conclude that the dependence of $M_{\rm crit}$ on $\rho_{\rm c}$ is quite
weak. An increase in $\rho_{\rm c}$ by a factor of three leads to at most a
$\sim 10$\% increase in the value of $M_{\rm crit}$, whereas a ten-fold
decrease in $\rho_{\rm c}$ decreases the critical mass by at most 30\%. We
interpret the insensitivity of the critical core mass to the value of
$\rho_{\rm c}$ as being due to a near cancellation of two opposing effects. On
the one hand, a denser core of a given mass commands a higher gravity, pulling
the surrounding gas closer in and thus increasing the chance of the
gravitational collapse of the atmosphere. On the other hand, a denser core
also implies a higher luminosity (under the assumption of a fixed cooling
time), since $r_{\rm c} \propto \rho_{\rm c}^{-1/3}$, which makes the collapse
less likely as discussed in previous sections of the paper.

\subsection{Dependence on collapse metallicity}\label{sec:z}

\begin{figure}
\psfig{file=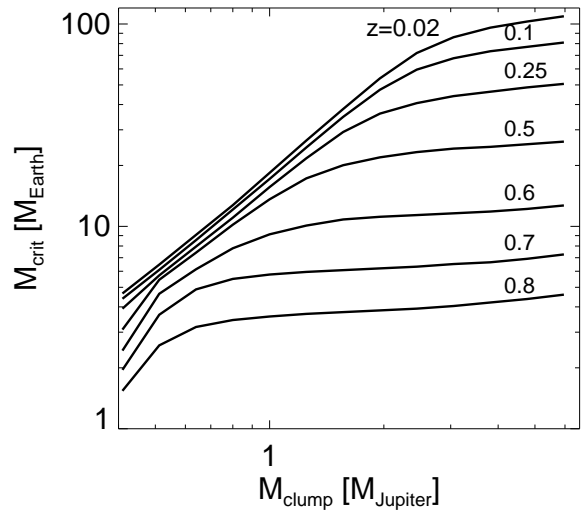,width=0.5\textwidth,angle=0}
\caption{Sensitivity of $M_{\rm crit}$ to the critical metallicity that
  triggers atmosphere collapse. Note that for any clump mass, collapse of the
atmosphere occurs for smaller critical masses as the metallicity of the gas is
increased.}
\label{fig:z_crit}
\end{figure}

A crucial argument was made in \S \ref{sec:accretion}: the collapse of a {\it
  metal-dominated} atmosphere does not in general signal the beginning of a
runaway {\it gas} accretion onto the core. Moreover, we argued that $z\sim0.5$ (and not higher) is
the critical metallicity for understanding the collapse of the entire system.
This model assumption is in variance to
\cite{HoriIkoma11} who suggested that critical core mass for core accretion
instability can be greatly reduced for strongly metal-polluted gas, so that
$M_{\rm crit}$ is as small as $\sim (0.2-1) \mearth$ for $z\simgt 0.8$.

We test the sensitivity of our results to the metallicity $z_{\rm
crit}$ of the atmosphere by varying the value of $z_{\rm crit}$ over a
broad range.  This also allows us to explore examples of systems for
which the atmosphere should collapse, but the clump as a whole should
not, as well as systems for which the instability of the atmosphere
can drive the collapse of the entire clump. Figure \ref{fig:z_crit}
shows a number of calculations identical to those done in \S
\ref{sec:param}, except just one particular core cooling time is used
($t_{\rm kh} = 10^6$~yr) and the metallicity is varied from $z_{\rm
crit}=0.02$ to $z_{\rm crit}=0.8$, as labelled on the figure.

Figure \ref{fig:z_crit} confirms that higher metallicity gas is more
susceptible to gravitational collapse (since $M_{\rm crit}$ is lower for
higher values of $z_{\rm crit}$). In this sense our results agree
qualitatively with that of \cite{HoriIkoma11}.  However, our results also show that the
collapse of a metal-dominated atmosphere does not obviously cause the collapse of the
rest of the gas clump, which is assumed to have ``normal'' metallicity, e.g.,
that of the protoplanetary disc. To see this, consider the evolution of
a growing, low-mass core, say $M_{\rm c}\simlt $ a few
$\mearth$. Only a highly metal-polluted layer, $1-z\ll 1$, could become
gravitationally unstable around such a low-mass core. For definitiveness,
consider $M_{\rm clump} = 2 M_J$ and $z=0.8$ case, for which $M_{\rm crit}
\approx 4 \mearth$. If the mass of the core roughly doubles when the whole
metal polluted layer collapses onto the core (because such a collapse occurs
when the mass of the metal-rich atmosphere is comparable to that of the core),
we would have an $\sim 8 \mearth$ core after the collapse. This is
insufficient to induce collapse of the gas beyond the metal polluted
layer. For the particular example at hand, a core mass of $\sim 55 \mearth$
is needed to induce the further collapse of an atmosphere at $z \sim 0.02$.
A new metal-polluted layer would have to build up, and that
would collapse at $z\sim 0.67$ or so for $M_{\rm crit} \approx 8 \mearth$.

Collapse of the entire clump, however, seems to become possible if the
instability in the atmosphere occurs at $z\sim 0.5$. For example, the critical
core mass for such a "moderately" polluted atmosphere is $M_{\rm crit} \approx
22 \mearth$. If $M_{\rm c}$ does approximately double after this collapse,
then even a normal metallicity gas, $z=0.02$, becomes unstable to $M_{\rm
  crit} \sim 50\mearth$. Collapse of a $z\sim 0.5$ atmospheric layer can thus
potentially start the collapse of the gas fragment itself. Earlier collapses
of metal-dominated layers should be interpreted as {\it core accretion},
albeit delayed and with some hydrogen rather than a true beginning of the gas
accretion runaway phase.

\subsection{Dependence on the metallicity-opacity coupling}\label{sec:opacity}

\begin{figure}
\psfig{file=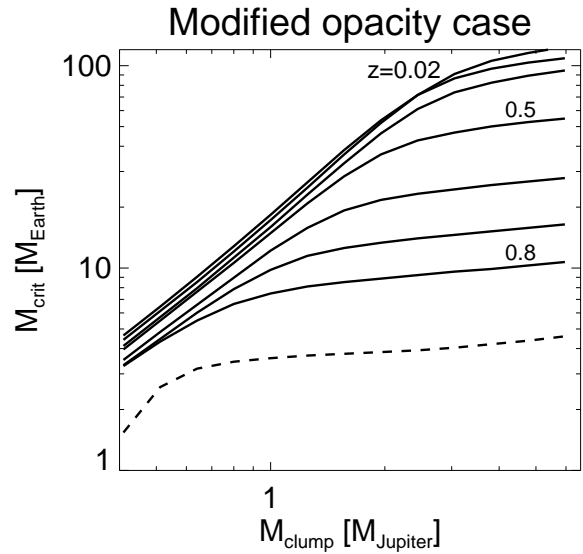,width=0.5\textwidth,angle=0}
\caption{Same as Figure \ref{fig:z_crit} but assuming that opacity in the
  metal-polluted layer is proportional to the metallicity of the gas (see text
  for more detail). The dashed curve is the $z=0.8$ curve from Figure
  \ref{fig:z_crit}, shown here for comparison. The opacity does have an effect
  on the critical core mass required to induce collapse.  However, this effect
  is limited to about a factor of two for the high-metallicity case.}
\label{fig:opacity}
\end{figure}

We have used the opacity of \cite{ZhuEtal09} in this paper up to now, which
includes dust, molecular, atomic, and electron scattering opacities in a
tabulated format. Unfortunately, these tables assume the single (Solar)
metallicity value, $z=0.02$. The grey opacity can be much higher for a higher
metallicity gas. For example, for dust opacity, a simple approximation is to assume that the opacity is directly proportional to the gas metallicity \citep[although
  changes in opacity due to grain growth could render this approximation
  questionable][]{HB10}.

 Clearly, $M_{\rm crit}$ does depend on the opacity that is used, and it is important
 to at least estimate how strongly.  Following grain growth and chemistry in
 detail is beyond the scope of this paper. Instead, we adopt the simple approach and assume that the grey opacity is directly
 proportional $z$. In particular, we multiply all the opacity sources from
 \cite{ZhuEtal09} by the factor $z/0.02$, except electron scattering, which
 is only important at very high $T$. 

As done in Figure \ref{fig:z_crit}, Figure \ref{fig:opacity}  explores the
 model $t_{\rm kh} = 10^6$~yr as a function of the metallicity of
the polluted layer, but now with the increased opacities, as described above
(solid curves). The same range of values for the layer's metallicity is used
in both figures. The dashed curve is the $z=0.8$ model from Figure
\ref{fig:z_crit} plotted again for comparison.

We see that increasing the opacity increases the critical mass for collapse of
the atmosphere, consistent with one's intuition and with the more detailed
study of \cite{HoriIkoma11}. Fortunately, this increase in $M_{\rm core}$ is
relatively small, e.g., only by a factor of $\sim 2$ for the highest
metallicity case, in which the opacities were increased by the factor of
$z/0.02=40$. This shows that opacity is indeed important in regulating the
critical core mass, but less so than the metallicity of the metal-polluted
layer. We draw a parallel here with previous work of \cite{IkomaEtal00} who
found that the critical core mass for atmosphere collapse in the CA theory
scales as a very weak power-law of the opacity: $M_{\rm crit} \propto k_{\rm
  gr}^s$, where $s=0.2-0.3$. Value of $s=0.2$, in particular, is consistent
with the increase of $M_{\rm crit}$ by a factor of 2 as opacity increases by a
factor of 40. The physical reason for this behavior appears to be the
  fact that most of the atmosphere's mass is located rather close to the core,
  where gas densities are very high (see e.g., figure 2), so that the
  convective rather than the radiative energy flux dominates.

\subsection{Dependence on the age of the fragment}\label{sec:t_age}

\begin{figure}
\psfig{file=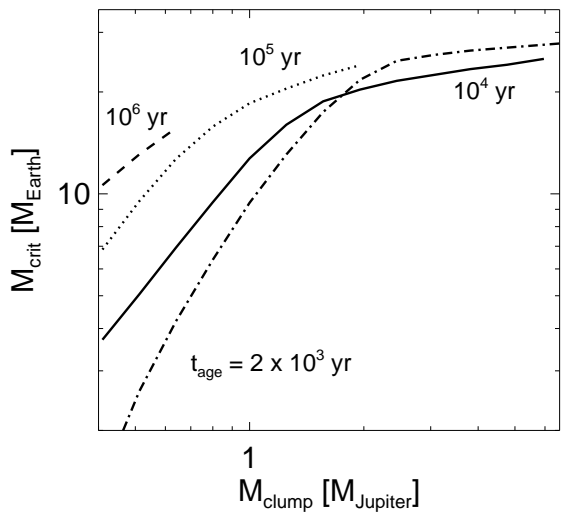,width=0.5\textwidth,angle=0}
\caption{Same as figure \ref{fig:z_crit} but assuming four different gas
  fragment ages, as marked on the figure next to the corresponding curves.
  The age of the clump has a very strong effect on the critical core mass, as
  the central properties of the clump vary significantly with time.  Three of
  the curves do not cover the full range of clump masses because the central
  temperatures exceed the dust evaporation temperature for more massive clumps
  (at the given time). }
\label{fig:age}
\end{figure}

Finally, our standard value for the age of the gas clump was set to $t_{\rm
  age}=10^4$ yr so far. This is a reasonable estimate for the gas fragment migration
time scale and the dust sedimentation time scale within the clump. However, a gas
fragment's thermal evolution may be affected by irradiation from the
surrounding disc and the parent star \citep[e.g.,][]{VazanHelled12}, and the
migration time may vary depending on the disc properties
\citep[e.g.,][]{Nayakshin10c}. Hence it is prudent to ask how our results
depend on the age of the clump. 

Figure \ref{fig:age} presents the critical core mass as a function of gas
clump's mass, $M_{\rm clump}$, for just one value of $t_{\rm kh} =
10^6$~yr, but for four different values of $t_{\rm age}$, from $2\times
10^3$~yr to $10^6$ yr. The dominant effect seen in the figure is due to
the contraction (and corresponding heating) of the clump with time.
For example, clumps with a mass higher than $M_{\rm clump} \approx 0.6 M_J$ are
hotter than $1400$~K in the simple model of \cite{Nayakshin10c}, so that grain
sedimentation and core growth is not possible at this time. This is why the
dashed curve continues only to this mass; higher mass clumps of this age would
be too hot to continue accreting cores by grain sedimentation.

This figure, along with the standard parameter search shown in Figure
\ref{fig:param1}, demonstrates that the internal structure of the gas fragment
-- controlled primarily by the clump's mass {\it and} age -- changes the
critical mass $M_{\rm crit}$ most strongly. Future detailed investigations
coupling the gas clump with the disc are required.

\section{Discussion}

\subsection{Main results of the paper}\label{sec:main}

We used a series of 1D simple hydrostatic structure models to explore the
stability of heavy element-rich atmospheres surrounding cores that are
embedded deep within gaseous clumps.  The clumps are envisaged to have formed
through the fragmentation of gravitationally unstable protoplanetary discs,
while the cores are assumed to have formed through grain sedimentation.  Our
calculations show that the core's atmosphere, i.e., the gas immediately
surrounding the core, can become unstable and collapse for a range of fragment
and core masses, as well as core luminosities.  While the collapse of a core's
atmosphere in and of itself does not immediately imply that the entire clump
will be unstable, experiments with the 1D radiation hydrodynamical code
presented in \cite{Nayakshin10b,Nayakshin10c} did lead to the whole gas clump
collapsing on a dynamical time scale for a few test problems explored. Since
such a clump collapse is driven by H2 molecule dissociation and H atom
ionisation presenting very large energy sinks it seems reasonable to expect
the whole clump collapsing once core atmosphere/envelope becomes
unstable. This suggests that a disc fragment could form a gas giant planet via
a novel channel that is discovered here. Namely, traditionally giant planet
formation by gravitationally unstable discs is believed to occur via radiative
cooling and contraction of self-gravitating gas fragments \citep[see the
  review by][]{HelledEtal13a}. Dust plays a passive (albeit important) role in
this traditional picture by providing and regulating the opacity of the
fragment \citep[e.g.,][]{HB10}. In the picture developed in our paper,
however, the dust plays a dynamically important role. By accumulating into a
massive dense core in the centre, dust may provide a significant destabilising
effect onto the whole gas fragment. This newly discovered core-assisted gas
capture instability (CAGC) is closely linked to the well known Core Accretion
instability
\citep{MizunoEtal78,Harris78,Mizuno80,Stevenson82,PollackEtal96,Rafikov06}.

To enable a quantitative study of the CAGC, we made a number of model
  choices and simplifying assumptions. The gas fragment is assumed to be
  isolated from external influences and its structure is approximated by a
  very simple analytical model of \cite{Nayakshin10c}. The dense core has a
  fixed density and its luminosity is given by the binding energy divided by a
  cooling time, $t_{\rm kh}$, which is an important free parameter. The region
  closest to the core is metal polluted in our model since grains vaporise
  when $T> T_{\rm v} = 1400$~K. We argued that gas fragment collapse can be
  trigerred only if metallicity of the polluted layer is $z\simlt 0.5$ as
  opposed to the limit $1-z\ll 1$. Within these assumptions, our main
quantitative results are as follows:

\begin{enumerate}
\item The critical core mass, $M_{\rm crit}$, that triggers the collapse of a
  core's atmosphere is dependent most strongly on the central density and
    temperature of the fragment, as well as the luminosity of the core.  
    More luminous cores (which corresponds to shorter $t_{\rm kh}$) have more
    tenuous atmospheres, as expected, and hence brighter cores must be more
    massive to provoke the collapse (see fig. \ref{fig:param1}). As a
  numerical example, in a 2 M$_J$ clump of $10^4$ years of age from its birth,
  a core mass greater than $\sim 45 \mearth$ is required to cause collapse if
  the cooling time of the core is $10^4$ yr. ``Only'' $M_{\rm crit} \approx
  12.5 \mearth$ is however needed for a cooling time of $t_{\rm kh} = 10^7$
  yr.

\item Since protoplanetary discs seem to disappear on time scale of order
  a few Million years, the most interesting value of $t_{\rm kh}$ to explore
  is of order 1 Million years or less (see \S \ref{sec:di} below on this
  point). The critical core mass is then (fig. \ref{fig:param1}) between $\sim
  5$ and $\sim 50\mearth$, depending on properties of the gas fragment.

\item Less massive and younger gas fragments require smaller $M_{\rm crit}$ at
  the same $t_{\rm kh}$ (cf. figures \ref{fig:param1} and \ref{fig:age}).

\item Atmospheres of bright (short $t_{\rm kh}$) cores are convective, whereas
  atmospheres of dim (long $t_{\rm kh}$) cores are at least partially
  radiative. For our default opacity, the atmospheres become completely
  convective at cooling times of $\simlt 10^4$ yr. In this case, the critical
  core mass is a function of the central pressure and temperature in the
  clump. At a given clump age and a clump opacity law, these are a function of
  the total clump mass, so the critical core mass becomes insensitive to the
  core luminosity/cooling times in the limit of shortest $t_{\rm kh}$ (cf. the
  upper enveloping (red) curve in figure \ref{fig:param1}).

\item $M_{\rm crit}$ is very weakly dependent on the core density and opacity
  in the envelope (\S\S 5.1 and 5.3).

\item The collapse of gas fragments to much higher proto-planetary densities
  via CAGC is strongly dependent on the availability of heavy elements, as
  larger critical core masses can only be reached for high initial
  metallicity.  {\it Therefore, based on these results, a strong positive
    correlation between frequency of giant planets and metallicity may be
    expected with CAGC}.  Low metallicity environments may be dominated by
  clumps that undergo collapse through H$_2$ dissociation.

\end{enumerate}

\subsection{Implications for the disc instability model}\label{sec:di}

In the disc instability paradigm
\citep[e.g.,][]{Kuiper51b,Boss97,DurisenEtal07}, gas fragments of a few
Jupiter masses are born in the outer cold and massive disc due to self-gravity
instability. These fragments then cool and eventually collapse via hydrogen
molecule dissociation in the fragment \citep{BodenheimerEtal80}. Rapid inward
migration of clumps
\citep[e.g.,][]{VB06,ChaNayakshin11a,BaruteauEtal11,MichaelEtal11,ZhuEtal12a}
may however lead to their destruction due to tides or irradiation
\citep{CameronEtal82,VazanHelled12} or tidal disruption
\citep{BoleyEtal10,Nayakshin10c}. It is hence only those clumps that
contracted before they were disrupted that may leave behind giant planets.

The newly found CAGC instability is an additional channel via which gas
fragments may collapse. Dust growth and sedimentation into a core can be
faster than the time taken by the fragments to cool to H$_2$ dissociation,
especially for fragments less massive than a few Jupiter masses
\citep{HS08,HB10,Nayakshin10a,Nayakshin10c}. Furthermore, such fragments are
also most vulnerable to external irradiation which slows down theor cooling
even further \citep{VazanHelled12}. Therefore, CAGC may be the dominant
channel through which such relatively low-mass fragments collapse to become
giant planets.

We caution however that much more work is needed to test these ideas. Adding a
self-consistent evolutionary model for the internal structure of a fragment
interacting with the disc and the protostar and migrating inward is
essential. In addition, there are several physical processes such as
turbulence and magnetic fields which are not included in this work. These
processes could affect the formation of clumps by gravitational instability
and should be included in future disc and planet formation models. Turbulence
could also slow down formation of cores inside the fragments.

\subsection{Implications for observations of giant planets}\label{sec:obs}

\begin{enumerate}

\item Presence of a massive core. Based on the older variants of disc
  instability models \citep[e.g.,][]{Boss97} that do not include planet
  migration and tidal stripping of the fragments, it is tempting to use the
  presence of a core in a gas giant as {\it prima facie} evidence of formation
  by core accretion plus gas capture, as core accretion {\it requires} the
  formation of a core \citep[e.g.,][]{Mizuno80,Stevenson82,PollackEtal96},
  while fragmentation by disc instability does not.  However, as emphasized
  here, core formation could occur nonetheless through grain sedimentation
  (e.g., Boss 1997, 1998; Helled \& Schubert 2008; Helled et al.~2008;
  Nayakshin 2010; Boley et al. 2010, 2011; Forgan \& Rice 2013), and more
  importantly, if it does, can promote the collapse of the fragment  if
    the core's mass exceeds $M_{\rm crit}$.  Sedimentation can hence erase or
  severely confuse using internal structure as evidence for a given formation
  mode, particularly if the formation of a heavy element core in a fragment
  can also lead to variations in final bulk composition, as discussed next.

 Since the presence of a core can be explained by both formation scenarios
  it becomes harder to discriminate among the two models based on the core's
  mass. While it is still widely believed that small cores are more consistent
  with disc instability while massive cores are a natural outcome of core
  accretion, we found in this paper that reality may be much more
  complex. Even the extreme cases (i.e., no core at all versus a very massive
  core) do not implicate the formation process directly. This is due to the
  fact that very massive cores can also form in the disc instability picture
  if enrichment from birth and/or planetesimal capture occur, while no or very
  small cores are not inconsistent with core accretion if core erosion takes
  place. We therefore suggest once more, that core mass should not be used as
  a proof for a given formation model.  In addition, one has to keep in mind
  that core masses are inferred only indirectly from models of their interior
  stucture \citep[e.g.,][]{MillerFortney11,PodolakHelled12}. For extrasolar
  giant planets it is not currently possible to distinguish between the total
  heavy element enrichment and high core masses \citep[e.g.,][]{HelledEtal13a}

\item Planet composition. Planets born by disc instability are often assumed
  to have the same composition as their host stars.  However, several
  processes can occur that could potentially alter this \citep[e.g., see
    review by][]{HelledEtal13a}.  As discussed above, the formation of a core
  in a clump must simultaneously deplete the clump's envelope of heavy
  elements.  If core-assisted gas capture has occurred prior to the loss of
  the clump's envelope during disc migration, then any gas that is lost to the
  disc through tides would be preferentially depleted in heavy elements.  Any
  remaining planet could thus appear to be significantly enriched in heavy
  elements.

\item Giant planet frequency of occurence in the inner fraction of an AU was
  found to be a strongly increasing function of metallicity of the host star
  \citep{FischerValenti05}. This has been traditionally interpreted as
  evidence for formation of these planets via Core Accretion picture. Our
  results however show that metal rich fragments born by the disc instability
  are more likely to collapse via CAGC, and it is thus possible that the
  observed correlation could be recovered by this model as well (cf. \S
  \ref{sec:param} and point (vi) in \S \ref{sec:main}). This result is
  potentially exciting, but further work is required to better understand the
dependence of CAGC on metallicity. On one hand, high heavy element abundances
implies higher opacities and longer overall clump cooling times and perhaps
larger cores \citep{Nayakshin10a}.  On the other hand, high initial abundances
could prompt prodigious grain growth. While this  should promote a faster
  core growth, it could also actually decrease the overall cooling time of the
  clump \citep[e.g.,][]{HB11}.  The uncertainty in opacity due to grain growth
  is however also a concern for the CA model \citep[e.g.,][]{Movshovitz2010}.

Despite these uncertainties, it seems unlikely for CAGC to be important
  for clumps more massive than a few to $\sim 10$ Jupiter masses, because such
  clumps are unable to form massive solid cores quicker than they heat up to
  vaporise their grains \citep{HS08,Nayakshin10a}. Therefore, even under most
  optimistic assumptions about the CAGC contribution to forming giant gas
  planets, it seems highly unlikely that CAGC could explain planet-metallicity
  correlation for planets more massive than $\sim 10$ Jupiter masses.

\end{enumerate}

\subsection{Implications for observations of sub-giant
    planets}\label{sec:hot} 

``Modern'' versions of the disc instability model, e.g., those that
  include core formation, fragment migration and disruption
  \citep[e.g.,][]{BoleyEtal10,Nayakshin10c} are multi-outcome valued due to a
  large number of physical processes operating. Due to these processes,
  results obtained here may have important implications for close in ``hot''
  (planet-star separation $a\sim 0.03 0.3$ AU) sub-giant planets as well.

In the context of such modern models, these close-in planets could have formed
by the same process of grain sedimentation within the gas fragments, but in
one of the following two distinctly different scenarios:
\begin{enumerate}
\item Hot rocky super-Earths or hot-neptune planets of mass $M_p$ are
  assembled by grain sedimentation inside gas fragments {\it that are
    disrupted before H$_2$ dissociation} while $M_p < M_{\rm crit}$. If $M_p
  \ll M_{\rm crit}$, only a tiny gas atmosphere would be retained after the
  disruption. If $M_p \simlt M_{\rm crit}$, a more massive atmosphere is
  retained, so that the planet may be neptune-like in terms of bulk properties
  and composition. Tidal disruption of pre-H$_2$ dissociation fragments can
  occur at distances $R\simgt 1$ AU to tens of AU, depending on the internal
  state of the fragments. For reference, tidal density around the Sun is
  $\approx 10^{-7}$~g~cm$^{-3}$. Cores are much denser, so they survive this
  disruption \citep{BoleyEtal10}. If they migrate inward via type I migration
  to $a\sim 0.1$~AU then they could potentially explain the population of
  close-in sub-giant exoplanets.

\item In the other scenario, H$_2$ dissociation and fragment collapse do take
  place before the clump is tidally disrupted. This could be either because
  $M_p \geq M_{\rm crit}$ (collapse via CAGC instability studied here) or
  because the fragment contracted sufficiently strongly for H$_2$ dissociation
  to take place \citep[as in the classical picture of,
    e.g.,][]{BodenheimerEtal80}, when $M_p < M_{\rm crit}$. The key point here
  is that such a collapse still does not necessarily implies the survival of
  the giant planet for astrophysically interesting time scales. For this to be
  the case, the planet must stop migrating towards the star. In the opposite
  case, the whole giant planet may be completely lost when it is swallowed by
  the parent star, or, alternatively -- if the planet contracts slower than it
  migrates inward -- it may be tidally disrupted in the ``hot region''
  \citep{Nayakshin11b}, $a\sim 0.1$ AU. In this last case just the core is
  left behind. In contrast to the picture discussed in (i) above, the
  migration of the planet in this case is via type II until the gas fragment
  is disrupted. \cite{NayakshinLodato12} showed that this close-in tidal
  disruption process reproduces some close in features of FU-Ori outbursts
  seen around some young stars \citep{HK96}.

\end{enumerate}

\section{Summary}

We showed that gas fragments born by disc instability in the outer cold
protoplanetary disc may collapse in a novel way when a massive dense core
composed of heavy elements builds up in the centre of the fragments.  The new
Core-Assisted Gas Capture (CAGC) instability may allow fragment collapse to
higher densities faster than by the traditional cooling and contraction route
\citep{BodenheimerEtal80}, especially for gas clumps less massive than a few
Jupiter masses. This may enhance survivability of rapidly migrating gas
fragments as giant planets. While significantly improved and expanded
follow-up calculations are needed, this new mode of giant planet formation may
show a positive correlation with host star's metallicity.

\section{Acknowledgments}

We thank A. Kovetz for providing us with the table EOS for hydrogen/metals
mix.  Theoretical astrophysics research in Leicester is supported by an STFC
grant. ACB's support is provided by The University of British Columbia.


\begin{thebibliography}{63}
\expandafter\ifx\csname natexlab\endcsname\relax\def\natexlab#1{#1}\fi

\bibitem[{Baruteau} et~al.(2011){Baruteau}, {Meru} \&
  {Paardekooper}]{BaruteauEtal11}
{Baruteau} C., {Meru} F., {Paardekooper} S.-J., 2011, \mnras, 416, 1971

\bibitem[{Bell} \& {Lin}(1994)]{Bell94}
{Bell} K.~R., {Lin} D.~N.~C., 1994, \apj, 427, 987

\bibitem[{Bodenheimer} et~al.(1980){Bodenheimer}, {Grossman}, {Decampli},
  {Marcy} \& {Pollack}]{BodenheimerEtal80}
{Bodenheimer} P., {Grossman} A.~S., {Decampli} W.~M., {Marcy} G., {Pollack}
  J.~B., 1980, \icarus, 41, 293

\bibitem[{Boley} et~al.(2010){Boley}, {Hayfield}, {Mayer} \&
  {Durisen}]{BoleyEtal10}
{Boley} A.~C., {Hayfield} T., {Mayer} L., {Durisen} R.~H., 2010, Icarus, 207,
  509

\bibitem[{Boley} et~al.(2011){Boley}, {Helled} \& {Payne}]{BoleyEtal11a}
{Boley} A.~C., {Helled} R., {Payne} M.~J., 2011, ArXiv e-prints

\bibitem[{Boss}(1997)]{Boss97}
{Boss} A.~P., 1997, Science, 276, 1836

\bibitem[{Boss}(1998)]{Boss98}
{Boss} A.~P., 1998, \apj, 503, 923

\bibitem[{Cameron} et~al.(1982){Cameron}, {Decampli} \&
  {Bodenheimer}]{CameronEtal82}
{Cameron} A.~G.~W., {Decampli} W.~M., {Bodenheimer} P., 1982, Icarus, 49, 298

\bibitem[{Cha} \& {Nayakshin}(2011)]{ChaNayakshin11a}
{Cha} S.-H., {Nayakshin} S., 2011, \mnras, 415, 3319

\bibitem[{Cossins} et~al.(2009){Cossins}, {Lodato} \& {Clarke}]{CossinsEtal09}
{Cossins} P., {Lodato} G., {Clarke} C.~J., 2009, \mnras, 393, 1157

\bibitem[{Durisen} et~al.(2007){Durisen}, {Boss}, {Mayer}, {Nelson}, {Quinn} \&
  {Rice}]{DurisenEtal07}
{Durisen} R.~H., {Boss} A.~P., {Mayer} L., {Nelson} A.~F., {Quinn} T., {Rice}
  W.~K.~M., 2007, Protostars and Planets V,  607--622

\bibitem[{Fischer} \& {Valenti}(2005)]{FischerValenti05}
{Fischer} D.~A., {Valenti} J., 2005, \apj, 622, 1102

\bibitem[{Forgan} \& {Rice}(2011)]{ForganRice11}
{Forgan} D., {Rice} K., 2011, \mnras, 417, 1928

\bibitem[{Forgan} \& {Rice}(2013)]{ForganRice13}
{Forgan} D., {Rice} K., 2013, \mnras, 430, 2082

\bibitem[{Gammie}(2001)]{Gammie01}
{Gammie} C.~F., 2001, \apj, 553, 174

\bibitem[{Guillot}(2005)]{Guillot05}
{Guillot} T., 2005, Annual Review of Earth and Planetary Sciences, 33, 493

\bibitem[{Harris}(1978)]{Harris78}
{Harris} A.~W., 1978, in { Lunar and Planetary Institute Science Conference
  Abstracts\/}, vol.~9 of { Lunar and Planetary Institute Science Conference
  Abstracts\/},  459--461

\bibitem[{Hartmann} \& {Kenyon}(1996)]{HK96}
{Hartmann} L., {Kenyon} S.~J., 1996, \araa, 34, 207

\bibitem[{Hartmann} et~al.(2011){Hartmann}, {Zhu} \& {Calvet}]{HartmannEtal11}
{Hartmann} L., {Zhu} Z., {Calvet} N., 2011, ArXiv e-prints

\bibitem[{Helled} \& {Bodenheimer}(2010)]{HB10}
{Helled} R., {Bodenheimer} P., 2010, \icarus, 207, 503

\bibitem[{Helled} \& {Bodenheimer}(2011)]{HB11}
{Helled} R., {Bodenheimer} P., 2011, \icarus, 211, 939

\bibitem[{Helled} et~al.(2013){Helled}, {Bodenheimer}, {Podolak}
  et~al.]{HelledEtal13a}
{Helled} R., {Bodenheimer} P., {Podolak} M., et~al., 2013, ArXiv e-prints

\bibitem[{Helled} et~al.(2006){Helled}, {Podolak} \& {Kovetz}]{HelledEtal06}
{Helled} R., {Podolak} M., {Kovetz} A., 2006, \icarus, 185, 64

\bibitem[{Helled} et~al.(2008){Helled}, {Podolak} \& {Kovetz}]{HelledEtal08}
{Helled} R., {Podolak} M., {Kovetz} A., 2008, Icarus, 195, 863

\bibitem[{Helled} \& {Schubert}(2008)]{HS08}
{Helled} R., {Schubert} G., 2008, Icarus, 198, 156

\bibitem[{Hori} \& {Ikoma}(2011)]{HoriIkoma11}
{Hori} Y., {Ikoma} M., 2011, \mnras, 416, 1419

\bibitem[{Ikoma} et~al.(2000){Ikoma}, {Nakazawa} \& {Emori}]{IkomaEtal00}
{Ikoma} M., {Nakazawa} K., {Emori} H., 2000, \apj, 537, 1013

\bibitem[{Kratter} et~al.(2010){Kratter}, {Murray-Clay} \&
  {Youdin}]{KratterEtal10}
{Kratter} K.~M., {Murray-Clay} R.~A., {Youdin} A.~N., 2010, \apj, 710, 1375

\bibitem[{Kuiper}(1951)]{Kuiper51b}
{Kuiper} G.~P., 1951, Proceedings of the National Academy of Science, 37, 1

\bibitem[{Larson}(1969)]{Larson69}
{Larson} R.~B., 1969, \mnras, 145, 271

\bibitem[{Marley} et~al.(2007){Marley}, {Fortney}, {Hubickyj}, {Bodenheimer} \&
  {Lissauer}]{MarleyEtal07}
{Marley} M.~S., {Fortney} J.~J., {Hubickyj} O., {Bodenheimer} P., {Lissauer}
  J.~J., 2007, \apj, 655, 541

\bibitem[{McCrea} \& {Williams}(1965)]{McCreaWilliams65}
{McCrea} W.~H., {Williams} I.~P., 1965, Royal Society of London Proceedings
  Series A, 287, 143

\bibitem[{Michael} et~al.(2011){Michael}, {Durisen} \& {Boley}]{MichaelEtal11}
{Michael} S., {Durisen} R.~H., {Boley} A.~C., 2011, \apjl, 737, L42+

\bibitem[{Miller} \& {Fortney}(2011)]{MillerFortney11}
{Miller} N., {Fortney} J.~J., 2011, \apjl, 736, L29

\bibitem[{Mizuno}(1980)]{Mizuno80}
{Mizuno} H., 1980, Progress of Theoretical Physics, 64, 544

\bibitem[{Mizuno} et~al.(1978){Mizuno}, {Nakazawa} \& {Hayashi}]{MizunoEtal78}
{Mizuno} H., {Nakazawa} K., {Hayashi} C., 1978, Progress of Theoretical
  Physics, 60, 699

\bibitem[{Mordasini}(2013)]{Mordasini13}
{Mordasini} C., 2013, ArXiv e-prints

\bibitem[{More} et~al.(1988){More}, {Warren}, {Young} \&
  {Zimmerman}]{MoreEtal98}
{More} R.~M., {Warren} K.~H., {Young} D.~A., {Zimmerman} G.~B., 1988, Physics
  of Fluids, 31, 3059

\bibitem[{Movshovitz} et~al.(2010){Movshovitz}, {Bodenheimer}, {Podolak} \&
  {Lissauer}]{Movshovitz2010}
{Movshovitz} N., {Bodenheimer} P., {Podolak} M., {Lissauer} J.~J., 2010,
  \icarus, 209, 616

\bibitem[{Nayakshin}(2010{\natexlab{a}})]{Nayakshin10c}
{Nayakshin} S., 2010{\natexlab{a}}, \mnras, 408, L36

\bibitem[{Nayakshin}(2010{\natexlab{b}})]{Nayakshin10a}
{Nayakshin} S., 2010{\natexlab{b}}, \mnras, 408, 2381

\bibitem[{Nayakshin}(2011{\natexlab{a}})]{Nayakshin10b}
{Nayakshin} S., 2011{\natexlab{a}}, \mnras, 413, 1462

\bibitem[{Nayakshin}(2011{\natexlab{b}})]{Nayakshin11b}
{Nayakshin} S., 2011{\natexlab{b}}, \mnras, 416, 2974

\bibitem[{Nayakshin} \& {Cha}(2013)]{NayakshinCha13}
{Nayakshin} S., {Cha} S.-H., 2013, \mnras

\bibitem[{Nayakshin} \& {Lodato}(2012)]{NayakshinLodato12}
{Nayakshin} S., {Lodato} G., 2012, \mnras, 426, 70

\bibitem[{Perri} \& {Cameron}(1974)]{PerriCameron74}
{Perri} F., {Cameron} A.~G.~W., 1974, \icarus, 22, 416

\bibitem[{Podolak} \& {Helled}(2012)]{PodolakHelled12}
{Podolak} M., {Helled} R., 2012, \apjl, 759, L32

\bibitem[{Podolak} et~al.(1988){Podolak}, {Pollack} \&
  {Reynolds}]{PodolakEtal88}
{Podolak} M., {Pollack} J.~B., {Reynolds} R.~T., 1988, \icarus, 73, 163

\bibitem[{Pollack} et~al.(1996){Pollack}, {Hubickyj}, {Bodenheimer},
  {Lissauer}, {Podolak} \& {Greenzweig}]{PollackEtal96}
{Pollack} J.~B., {Hubickyj} O., {Bodenheimer} P., {Lissauer} J.~J., {Podolak}
  M., {Greenzweig} Y., 1996, Icarus, 124, 62

\bibitem[{Prialnik} \& {Livio}(1985)]{PrialnikLivio85}
{Prialnik} D., {Livio} M., 1985, \mnras, 216, 37

\bibitem[{Rafikov}(2006)]{Rafikov06}
{Rafikov} R.~R., 2006, \apj, 648, 666

\bibitem[{Sasaki}(1989)]{Sasaki89}
{Sasaki} S., 1989, \aap, 215, 177

\bibitem[{Saumon} et~al.(1995){Saumon}, {Chabrier} \& {van Horn}]{SaumonEtal95}
{Saumon} D., {Chabrier} G., {van Horn} H.~M., 1995, \apjs, 99, 713

\bibitem[{Stamatellos} \& {Whitworth}(2008)]{SW08}
{Stamatellos} D., {Whitworth} A.~P., 2008, \aap, 480, 879

\bibitem[{Stevenson}(1982)]{Stevenson82}
{Stevenson} D.~J., 1982, P\&SS, 30, 755

\bibitem[{Toomre}(1964)]{Toomre64}
{Toomre} A., 1964, \apj, 139, 1217

\bibitem[{Vazan} \& {Helled}(2012)]{VazanHelled12}
{Vazan} A., {Helled} R., 2012, \apj, 756, 90

\bibitem[{Vazan} et~al.(2013){Vazan}, {Kovetz}, {Podolak} \&
  {Helled}]{VazanEtal13}
{Vazan} A., {Kovetz} A., {Podolak} M., {Helled} R., 2013, \mnras, 434, 3283

\bibitem[{Vorobyov} \& {Basu}(2005)]{VB05}
{Vorobyov} E.~I., {Basu} S., 2005, \apjl, 633, L137

\bibitem[{Vorobyov} \& {Basu}(2006)]{VB06}
{Vorobyov} E.~I., {Basu} S., 2006, \apj, 650, 956

\bibitem[{Wuchterl}(1993)]{Wuchterl93}
{Wuchterl} G., 1993, \icarus, 106, 323

\bibitem[{Zhu} et~al.(2009){Zhu}, {Hartmann} \& {Gammie}]{ZhuEtal09}
{Zhu} Z., {Hartmann} L., {Gammie} C., 2009, \apj, 694, 1045

\bibitem[{Zhu} et~al.(2012){Zhu}, {Hartmann}, {Nelson} \& {Gammie}]{ZhuEtal12a}
{Zhu} Z., {Hartmann} L., {Nelson} R.~P., {Gammie} C.~F., 2012, \apj, 746, 110

\end{thebibliography}

\label{lastpage}

\end{document}